# Efficient Dynamical Downscaling of General Circulation Models Using Continuous Data Assimilation

Srinivas Desamsetti[1], Hari Prasad Dasari[1], Sabique Langodan[1], Edriss S. Titi[2], Omar Knio[1] and Ibrahim Hoteit[1,*]

1. King Abdullah University of Science and Technology (KAUST), Physical Sciences and Engineering Division, Thuwal, Saudi Arabia

2. Department of Mathematics, Texas A&M University, College Station, USA

**\*Corresponding author:** Prof. Ibrahim Hoteit,
King Abdullah University of Science and Technology (KAUST),
Physical Science and Engineering Division,
Thuwal 23955-6900, Saudi Arabia.
E-mail: ibrahim.hoteit@kaust.edu.sa

**Abstract**

Continuous data assimilation (CDA) is successfully implemented for the first time for efficient dynamical downscaling of a global atmospheric reanalysis. A comparison of the performance of CDA with the standard grid and spectral nudging techniques for representing long- and short-scale features in the downscaled fields using the Weather Research and Forecast (WRF) model is further presented and analyzed. The WRF model is configured at $0.25° \times 0.25°$ horizontal resolution and is driven by $2.5° \times 2.5°$ initial and boundary conditions from NCEP/NCAR reanalysis fields. Downscaling experiments are performed over a one-month period in January, 2016.

The similarity metric is used to evaluate the performance of the downscaling methods for large (2000 km) and small (300 km) scales. Similarity results are compared for the outputs of the WRF model with different downscaling techniques, NCEP/NCAR reanalysis, and NCEP Final Analysis (FNL, available at $0.25° \times 0.25°$ horizontal resolution). Both spectral nudging and CDA describe better the small-scale features compared to grid nudging. The choice of the wave number is critical in spectral nudging; increasing the number of retained frequencies generally produced better small-scale features, but only up to a certain threshold after which its solution gradually became closer to grid nudging. CDA maintains the balance of the large- and small-scale features similar to that of the best simulation achieved by the best spectral nudging configuration, without the need of a spectral decomposition. The different downscaled atmospheric variables, including rainfall distribution, with CDA is most consistent with the observations. The Brier skill score values further indicate that the added value of CDA is distributed over the entire model domain. The overall results clearly suggest that CDA provides an efficient new approach for dynamical downscaling by maintaining better balance between the global model and the downscaled fields.

## 1. Introduction

Dynamical downscaling using Regional Atmospheric Models (RAMs) is a broadly recognized approach for resolving high resolution regional atmospheric features, e.g., Dickinson et al., (1989); Giorgi, (1990); Jacob and Podzun, (1997); Giorgi and Mearns, (1999); Meehl et al., (2007); Rinke and Dethloff, (2000); Dasari et al., (2010, 2014), Srinivas et al., (2015); Yesubabu et al., (2016), to cite but a few. It is widely implemented in wide-range of applications, such as local weather forecasts, regional climate change projections, air quality studies, energy applications, and numerous industrial applications (Jacob and Podzun, 1997; Meehl et al., 2007; Langodan et al., 2014, 2016).

Although dynamical downscaling has been proven efficient for producing high resolution information, the resulting outputs may still hold systematic and transient errors (von Storch et al., 2000; Leung and Gustafson, 2005; Steiner et al., 2006). Generally, the main sources of errors in dynamical downscaling result from either imperfect model physics, and from the dynamical downscaling method itself (Dickinson et al.,1989; Giorgi, 1990). The uncertainties related to the model, such as the size of the domain, vertical and horizontal resolutions, spin-up period, topography, and physical parameterizations, have been investigated in several previous studies, e.g., Davies, (1983); Giorgi and Mearns, (1999); Denis et al., (2002); Vincent and Hahmann, (2015).

To simulate fine-scale solutions, a balance between the RAM and the global down scaled fields should be maintained during the downscaling simulations by retaining the global large-scales and evolving the RAM to generate its own small-scale features; this has been considered as one of the most challenging problems in dynamical downscaling (Rockel et al., 2008; Vincent and Hahmann, 2015). To address this issue, the lateral boundary relaxation, or nudging, technique was introduced by Davis (1976). It basically consists of adding a nudging term to the predictive equation of the variable to be nudged. In this study, we investigate an innovative dynamical downscaling technique for RAMs, called Continuous Data Assimilation (CDA), and compare its performance with the state-of-the-art methods, the grid and spectral nudging techniques.

To capture the features of the driving large-scale fields throughout the domain, the grid (Stauffer and Seaman, 1990) and spectral (Waldron et al., 1996; von Storch, 1995; von Storch et al., 2000) nudging techniques have been proposed. Grid nudging is performed at every grid point of the domain, constraining with equal weights the whole spectrum of the atmospheric phenomena to the global fields (Stauffer and Seaman 1990). Spectral nudging aims at better maintaining the balance of the RAMs by only constraining the large-scale features while allowing the RAM to develop its local variability (Miguez-Macho et al., 2005). This technique allows the model to better represent the small-scale effects due to topography, land-sea contrast, and land-use distribution, and their interactions with the large-scale fields (Feser, 2006; Feser and von Storch, 2005; Rockel et al., 2008; Winterfeldt and Weisse, 2009; Vincent and Hahmann, 2015). However, the performance of spectral nudging strongly depends on the choice of the cut-off wave number, with no systematic way to set the value of this threshold other than conducting trial-and-error runs (von Storch et al., 2000; Liu et al., 2012). Closely related methods have been also proposed, replacing the large-scale fields of the RAM with the corresponding large-scale fields of the GCM at specified time intervals as in Kida et al. (1991) and Sasaki et al. (1995), or adding finer-scale perturbations to the large-scale GCM solution (Juang and Kanamitsu, 1994; Juang et al., 1997) within the spirit of ''anomaly models'' (Navarra and Miyakoda, 1988).

Continuous Data Assimilation (CDA) methods (Charney et al., 1969; Daley, 1991) can assimilate atmospheric observations into dynamical models during the integration time. Recently, significant progress has been made in the CDA approach by introducing a nudging term to the model equations to directly assimilate the observations (Henshaw et al., 2003; Korn, 2009; Olson and Titi, 2009; Hayden-, et al., 2011; Azouani et al., 2014; Bessaih et al., 2014; Altaf et al., 2017). The introduced nudging term constrains the model large-scale variability to available information, which is computed as a misfit between interpolants of the assimilated coarse grid information and fine grid model predictions. This new CDA method has been designed, implemented, and tested for different physical dynamical systems, including the Navier-Stokes equations, Rayleigh-Benard convection model and planetary geostrophic ocean circulation model (Azouani et al., 2014; Farhat et al., 2015, 2016a, b, c; Gesho et al., 2016; Altaf et al., 2017).

This study presents the first successful implementation of CDA for dynamical downscaling with a three-dimensional, non-hydrostatic regional circulation atmospheric model. We use the Advanced Research Weather Research and Forecasting (WRF-ARW; Skamarock et al., 2008) model version 3.9 developed by NCEP/NCAR for this purpose. We further evaluate and compare the results of the WRF simulations with grid nudging, spectral nudging, and CDA for resolving the large- and small-scale atmospheric features of a regional domain covering most of the African continent and the Middle East. Section 2 describes the model and the different downscaling techniques. Section 3 presents the evaluation method used for the analysis of the downscaled fields. The results are discussed in Section 4. Section 5 summarizes the main results of the study.

## 2. Model and nudging methods

The Advanced Research WRF (WRF-ARW) model version 3.9 developed by NCEP/NCAR (Skamarock et al., 2008) is used to conduct the dynamical downscaling experiments. The model configuration consists of a single domain covering the Africa continent and the Middle East with 25 km x 25 km (~0.25° × 0.25°) horizontal resolution and 35 vertical levels. Terrain elevation, land-use, and soil types were obtained from the United States Geological Survey (USGS) data available at arc 2' resolution. The simulations are performed over a one-month period starting from January 1, 2016. The model initial and 6- hourly boundary conditions are extracted from the NCEP/NCAR reanalysis data available at 2.5° × 2.5°.

Different model free runs were conducted without nudging (control run), and with the three downscaling methods: grid and spectral nudging, and CDA. All experiments were performed with the same WRF configuration. The grid and spectral nudging methods are implemented as in WRF-3.1.1 following Liu et al. (2012). CDA is implemented under the same conditions as spectral nudging. Nudging is performed every 6 hours over the entire simulation period and at every model grid point, with a nudging coefficient of 0.0003 $s^{-1}$ for all nudged variables (Stauffer and Seaman, 1990; Gomez and Miguez, 2017). The default spline interpolation operator in WRF model is used as the interpolant for the downscaling experiments with CDA. We also performed several sensitivity experiments using spectral nudging with different cut-off wave numbers (three, five, seventh, nine, eleven and thirteen) in both the x- and y-directions to investigate its sensitivity to the choice of this threshold; these are referred to as S33, S55, S77, S99, S1111, and S1313, respectively.

## 2.1 Downscaling techniques

Nudging or Newtonian relaxation is commonly used in RAMs to maintain consistency with the large-scale forcing fields while allowing mesoscale features to develop their own variability in the regional simulations (von Storch, 1995; von Storch et al., 2000; Hogrefe et al., 2004; Leung and Gustafson, 2005; Steiner et al., 2006). In this method, the model state is relaxed toward a reference (could be an observation or an analysis) state by adding an artificial tendency term, which is computed based on the difference between the observed and model predicted states. This section briefly summarizes the general formulations, and the differences, of the three tested downscaling methods. WRF is used in this study, and since its nudging implementation is based on Stauffer and Seaman (1990), we follow the same notations, in which nudging is defined as:

$$\frac{\partial \alpha}{\partial t} = F(\alpha, \vec{x}, t) + G_\alpha W(\vec{x}, t) \varepsilon(\vec{x})(\hat{\alpha}_0 - \alpha), \qquad \text{------------ (1)}$$

where the term $F(\alpha, \vec{x}, t)$ is the tendency predicted by the atmospheric model, $\vec{x}$ represents the spatial variables (x, y, z), $\alpha(\vec{x}, t)$ is a particular atmospheric state variable (to be nudged), and $\hat{\alpha}_0$ is the value toward which the state variable is nudged. $G_\alpha$ is the nudging factor that determines the relative magnitude of the tendency term in relation to the rest of the model processes included in $F(\alpha, \vec{x}, t)$. In WRF, its spatial and temporal variations are set by a time-dependent (four-dimensional) weighting function $W(\vec{x}, t) = W_{xy} W_z W_t$, in which $W_{xy}$ and $W_z$ are respectively the horizontal and vertical weighting functions defined based on a radius of influence and the distance from the observation. Similarly, the time weighting function $W_t$ depends on the model-relative time and the time of the observation (Stauffer and Seaman, 1990). The analysis quality factor $\varepsilon(\vec{x})$, typically varying between 0 and 1, depends on the distribution and quality of the nudging data. The nudged variables, α, include the zonal and meridional wind components, and the potential temperature (or the water vapor mixing ratio).

### 2.1.1 Grid nudging

Grid nudging in the WRF modeling system does not consider the quality of the analysis $\varepsilon(\vec{x})$ but includes a vertical weighting factor $V(z)$ with values ranging between 0 and 1. $V(z)$ is included to remove the impact of nudging near the surface, so that to allow the downscaling model to develop its own physics in the lower levels while nudging the circulation in the upper levels to the reference data (Stauffer and Seaman, 1990). The nudging equation in the WRF model is then expressed as

$$\frac{\partial \alpha}{\partial t} = F(\alpha, \vec{x}, t) + G_\alpha W(\vec{x}, t) V(z)(\hat{\alpha}_0 - \alpha), \qquad \text{----------- (2)}$$

The model prediction is therefore nudged towards a reference field, typically a coarse resolution global analysis, after it was interpolated to the RAM's grid. Eq. (2) is then applied assuming a perfect observation at every grid point.

### 2.1.2 Spectral nudging

Spectral nudging is implemented in a similar way to grid nudging, after applying a spectral filtering to the tendency term $(\hat{\alpha}_0 - \alpha)$, first in the x-direction and then in the y-direction. No filtering is applied in the vertical direction. The spectral nudging equation is then expressed as:

$$\frac{\partial \alpha}{\partial t} = F(\alpha, \vec{x}, t) + G_\alpha W(\vec{x}, t) V(z) \text{Filt}_{xy}[(\hat{\alpha}_0 - \alpha)], \qquad \text{------------- (3)}$$

where $\text{Filt}_{xy}$ represents a spectral filtering above a certain cut-off wave number. Filtering is then performed in three steps, as follows:

i) A Fast Fourier Transform (FFT) algorithm is first applied on each row of the tendency term $(\hat{\alpha}_0 - \alpha)$ to transform it to the spectral space.

ii) All wave numbers above a certain cut-off wavenumber in x-direction are set to zero.

iii) Then, using the inverse FFT, the remaining Fourier coefficients are transformed back to the spatial space.

The same procedure is then applied to each column of $(\hat{\alpha}_0 - \alpha)$ in the y-direction. The spectral filtering removes all spatial frequencies higher than the selected cut-off wave number, ensuring that nudging is only applied on the low wavelengths. Due to the orthogonality of the functions of the Fourier expansion, only the same spectral components of the physical space term $F(\alpha, \vec{x}, t)$ in Eq. (3) are affected by nudging.

**2.1.3 Continuous Data Assimilation (CDA)**

CDA exploits the fact that instabilities in turbulent flows occur at spatial large-scales and that spatial small-scales are stabilized by the viscous dissipation term in the Navier-Stokes equations (Currie, 1974). A rigorous mathematical framework was then developed showing that indeed the asymptotic, in time, behavior of the spatial large-scale of any solution to the Navier-Stokes equations determines in a unique fashion the asymptotic behavior of the full solution (Foias and Prodi, 1967; Jones and Titi, 1993; Cockburn et al., 1997). As such, the potential issue of solution multiplicity in numerical weather models (Weisse et al., 2000) would not manifest itself unless an extended setup is considered, involving for e.g. ensembles of uncertain trajectories, and/or stochastic parameterizations. Such a stochastic framework is yet to be covered by the theory of CDA.

To fix ideas, let us present some examples of the above results. Let us divide the physical domain $\Omega$ into disjoint subdomains $\Omega_j$, $j = 1, \cdots, N$, such that $diam(\Omega_j) \leq h$, and let us choose randomly points $\vec{x}_j \in \Omega_j$ for $j = 1, \cdots, N$. Suppose $\alpha(\vec{x}, t)$ and $\beta(\vec{x}, t)$ are two different solutions of the Navier-Stokes equations. If $\left(\alpha(\vec{x}_j,t) - \beta(\vec{x}_j,t)\right) \to 0$ as $t \to \infty$ for $j = 1, \cdots, N$, then $\left(\alpha(\vec{x},t) - \beta(\vec{x},t)\right) \to 0$ as $t \to \infty$ for all $\vec{x} \in \Omega$, provided $h \leq h_0$, where $h_0$ is a length scale that depends only on the Reynolds number, but is independent of the specific solutions. This means that the nodal values of the solutions at a spatial coarse scale of size $h$ determine the solution on the whole domain, asymptotically in time. Similarly, let us set $\bar{\varphi}_j \equiv \frac{1}{|\Omega_j|} \int_{\Omega_j} \varphi(\vec{x},t) dx$, the local volume average of the function $\varphi$, for $j = 1, \cdots, N$. If $\left(\bar{\alpha}_j(t) - \bar{\beta}_j(t)\right) \to 0$ as $t \to \infty$ for $j = 1, \cdots, N$, then $\left(\alpha(\vec{x},t) - \beta(\vec{x},t)\right) \to 0$ as $t \to \infty$ for all $\vec{x} \in \Omega$, provided $h \leq h_0$ as before. This means that the spatial coarse scale local volume averages at the domain of size $h$ determine the solution in a unique fashion, asymptotically in time.

Following the above notations, once can introduce interpolants approximation operators based on the model values or the local volume averages, respectively. Specifically, one can introduce, for example, the interpolant approximation operators by step functions as follows:

$$I_h(\varphi)(\vec{x}) = \sum_{j=1}^{N} \varphi(\vec{x}_j) \chi_{\Omega_j}(\vec{x}) \quad \text{or} \quad I_h(\varphi)(\vec{x}) = \sum_{j=1}^{N} \bar{\varphi}_j \chi_{\Omega_j}(\vec{x})$$

for the case of nodal values or the case of local value averages, respectively. Here, $\chi_{\Omega_j}(\vec{x}) = 1$ whenever $\vec{x} \in \Omega_j$; and $\chi_{\Omega_j}(\vec{x}) = 0$ whenever $\vec{x} \notin \Omega_j$ the characteristic function of the subdomain $\Omega_j$. The above interpolant operators $I_h(\varphi)$ are approximation functions of $\varphi$ at the spatial scale of size $h$.

Capitalizing on these results, Azouani et al. (2014) introduced a new approach for CDA that uses as nudging function the difference between coarse-scale interpolant of the spatial downscaled data and a coarse-scale interpolant of the model outputs. Note that $I_h(\varphi)$ are functions, and hence the nudging is not only at the grid points, but it is done adequately on a full neighborhood diameter size $h$ at each grid point. This sets constraints on the spatial large-scale flow of the model, which in turn forces the solution of the model to behave like the unknown reference solution that corresponds to the observational coarse mesh data. In particular, Azouani et al. (2014) were also able to show that the method is not sensitive to the model initial conditions, i.e., no matter how one initializes the model, its solution always converges, at an exponential rate in time, to the same unique reference solution that corresponds to the given coarse mesh data, provided the observational grid is fine enough depending on the Reynolds number. The CDA equation can be expressed as

$$\frac{\partial \alpha}{\partial t} = F(\alpha, \vec{x}, t) + G_\alpha W(\vec{x}, t) V(z)[I_h(\widehat{\alpha}_0) - I_h(\alpha)], \qquad \text{------------- (4)}$$

where $I_h$ is linear, coarse-mesh interpolation operator (Azouani et al., 2014). Specifically, $I_h$ is used to define a smooth function that (i) interpolates the data provided at the nodes of the coarser spatial mesh, and (ii) varies linearly between neighboring nodes.

Furthermore, stability analysis has enabled Mondaini and Titi (2018) and Ibdah et al., (2018) to establish uniform in time error estimates for the spatial discretization and the full discretization of the model, respectively, which makes its computational implementation reliable. Notably, it has also been observed that this CDA approach is equally applicable to other relevant dissipative systems, and that for certain systems it is sufficient to collect coarse mesh measurements of only part of the state variables (Farhat et al., 2016a,b,c; 2017). In this context, it has been shown rigorously (Farhat et al., 2016c) that for planetary scale geostrophic circulation models, the coarse mesh observations of the temperature are sufficient for determining and recovering the full reference solution, i.e., both the velocity and the temperature, as it has been asserted by Charney et al. (1969). In addition, this CDA approach has also been extended by Foias et al. (2016) to incorporate fully-discrete, in time and space, observations and to extract from these measurements statistical information concerning the corresponding unknown reference solution. Biswas et al. (2018) have recently extended this approach for recovering the probability distribution of the reference solutions from probability distribution of the observed measurements.

## 3. Evaluation method

To evaluate the results of the different downscaling techniques at different scales, we used the similarity concept proposed by von Storch et al. (2000) and implemented, for example, in Liu et al. (2012). The similarity index is measured based on a metric, $P(t, L)$, defined as

$$P(t, L) = 1 - \frac{\langle [\psi(t,L) - \psi^*(t,L)]^2 \rangle}{\langle \psi(t,L)^2 \rangle}, \quad \text{---------- (5)}$$

where $t$ is the model simulation time, $L$ the length scale of interest, $\psi(t, L)$ the input field (NCEP/NCAR data in this study) to the RAM, $\psi^*(t, L)$ the RAM (WRF) output field, $\langle \ \rangle$ denotes the 2D spatial-average over the domain. Every 6-hours, the similarity at different scales of interest are computed after obtaining $\psi(t, L)$ and $\psi^*(t, L)$, based on which the performances of the downscaling techniques are evaluated for both the large- and small-scales in opposing ways. A high similarity is desired for large-scales, as it suggests consistency between the large-scale features of the downscaled fields and the input fields (von Storch et al., 2000; Liu et al., 2012). In contrast, lower similarities are expected for the small-scale features in the RAM simulation as it develops its own variability that is not present in the global fields.

Following Liu et al. (2012), we compare the large- and small-scale features from NCEP and the RAM at the horizontal scales 2000 km and 300 km, respectively. To compute the similarity at these scales, we first interpolate NCEP/NCAR data (the input to WRF) to the same grid as the RAM (WRF, 25 km). The 25 km resolution grid cells in the modeling domain are next re-divided according to the scale of interest, in which the new cell includes several original grid cells. The representative values of the input, $\psi(t, L)$ and output, $\psi^*(t, L)$, fields are then computed for each new cell.

To evaluate the small-scale features in the downscaled WRF fields we introduce an independent dataset, NCEP Final Analysis (FNL) available at $0.25° \times 0.25°$ (which is of same grid resolution as our downscaled fields) as suggested by Liu et al. (2012). The idea is to compute the similarities between NCEP /NCAR and FNL for both large- and small-scales and use these as the reference to evaluate the similarities between the downscaled WRF and NCEP/NCAR fields at the small-scales. If FNL and NCEP/NCAR are consistent (i.e. have high similarity) at the large-scale, then the similarity between FNL and NCEP/NCAR at the small-scales could be used as criteria for a reasonable similarity of the WRF and NCEP/NCAR small-scale results. If not, the latter cannot be used directly and instead the difference of similarity between the large- and small-scales would be used to assess whether the change in similarity between the input and downscaled fields is reasonable (Liu et al., 2012).

To further assess quantitatively the solution of the nudging methods, we evaluated the Added Value (AV) of the downscaled fields with respect to the FNL observations compared to the model solution without nudging using the modified Brier Skill Score (BSS, von Storch and Zwiers, 1999; Feser et al., 2011; Li et al., 2016), defined as:

$$BSS = 1 - \frac{\sigma^2(\alpha_m, \alpha_o)}{\sigma^2(\alpha_c, \alpha_o)}, \text{ if } \sigma^2(\alpha_m, \alpha_o) < \sigma^2(\alpha_c, \alpha_o), \quad \text{----------} \quad (6)$$

$$BSS = \frac{\sigma^2(\alpha_c, \alpha_o)}{\sigma^2(\alpha_m, \alpha_o)} - 1, \text{ if } \sigma^2(\alpha_m, \alpha_o) > \sigma^2(\alpha_c, \alpha_o), \quad \text{----------} \quad (7)$$

where $\sigma$ is the root-mean-square-error (rmse) between two atmospheric variable solutions. $\alpha_o, \alpha_c,$ and $\alpha_m$ denote the FNL observations, the model control run without nudging, and the model solutions with the different nudging techniques (grid, spectral, CDA), respectively. Based on this definition, BSS varies between −1 and 1, with negative BSS suggesting the control run is more in agreement with the observations, and positive BSS indicating WRF with nudging has generated AV over the control run.

## 4. Results

This section assesses the downscaling performance of the CDA method and compares its results to those of the control, grid nudging, and spectral nudging solutions. The evaluation is performed based on surface pressure (PS), temperature (T850 and T500), the zonal (U) and meridional (V) wind components, and kinetic energy (KE850 and KE500), as a surrogate for wind speed, at the 850 and 500 hPa levels over the simulation period January 1–31, 2016. The similarity metrics $P(t, L)$ for these variables at large- and small-scales are first computed between the model and NCEP/NCAR, as described in Section 3, and their temporal mean and standard deviation are outlined in Table 1 and Table 2, respectively. The time evolutions of the similarity for large- and small-scales for temperature (T850 and T500), and kinetic energy (KE850 and KE500) are shown in Figures 1 and 2.

Both large- (Table 1) and small-scales (Table 2) of PS, T850, and T500 show a similar magnitude (> 0.9999) of similarity between CDA, grid, spectral, and no-nudging. In all experiments and over the whole downscaling period, the high similarity of T850, and T500 for the large-scale (Fig. 1a, d) suggest that all the downscaling techniques successfully capture the features of the driving large-scale fields. Although the differences in the magnitudes are relatively small in all the experiments, the similarity for temperature with NCEP/NCAR is relatively higher at small-scale for CDA and spectral nudging (Fig. 1b, e). Small temporal variations in the similarity for PS at both large- and small-scales indicate (not shown) a slight improvement in similarity at both scales with the spectral nudging and CDA compared to control run and grid nudging.

The similarity between the model and NCEP/NCAR at large-scale for KE850 and KE500 (Table 1) show (Figure 2) significant differences between the downscaling methods. All evaluated downscaling methods, grid, spectral, and CDA, suggest much higher means in similarity (> 0.9) at the large-scale than the control run without nudging (<0.47), outlining the ability of the downscaling techniques to retain the features of the global downscaled fields. In accordance with the discussion in Section 3, the similarity values (Table 2) are significantly lower for the small-scale for all three downscaled simulations. These lower similarities are expected due to the added variability produced by the RAMs (Liu et al., 2012).

Important differences were observed in the sensitivity runs of spectral nudging with different cut-off wave number. For instance, an increase in the retained spectral wave frequencies from S33 to S99 increased the similarity between grid and spectral nudging solution at both large- and small-scales. The differences between S77 and S99 were not significant, while the results of S1111 and S1313 quickly deteriorated (not shown). This confirms that spectral nudging should be implemented with a suitable number of waves, large enough to constrain the large-scales, but not too large to avoid the damping of the small-scales, as in grid nudging.

The high similarity between the NCEP/NCAR and FNL fields at the large-scales (Table 1) confirms the consistency between these two products. As discussed in Section 3, we further computed the similarity between NCEP/NCAR and FNL at small-scale (Table 2) to determine the consistency of lower similarities of the different downscaled fields at these scales. Although all the downscaling techniques provide a lower similarity at the small-scale (Fig. 1c, Fig. 1f, Fig. 2c, Fig. 2f) compared to the large-scale, CDA and S99 are consistent in producing comparable decrease in similarity to that of FNL. The close results of CDA to those obtained with the best simulation achieved using spectral nudging (S99) at both large- and small-scales indicate the advantage of CDA in producing robust downscaled fields without the need of any spectral decomposition.

As outlined in Table 2, on average and for both large and small scales, the differences in the similarity indices of the different downscaling techniques are in the order of $10^{-6}$, while their standard deviations are in the order of $10^{-7}$. To assess the significance of these small differences in the similarity indexes in term of the final downscaled solution, we compared the distributions of the differences between the WRF outputs (with the different downscaling techniques) and NCEP/NCAR at the large- and small- scales with those of the differences between FNL and NCEP/NCAR (Tables 3 and 4). The width of the distribution represents the variability added by the different downscaling techniques to global coarse fields (Liu et al., 2012). A small width reflects an over-nudging of the downscaled fields towards the driving global fields, while a larger width emphasizes the capability of the downscaling model to internally develop its small-scale variability that is missing in the driving coarse fields. The differences in the distribution width of the large- and small-scales for temperature and wind components at different levels using grid nudging are quite small compared to those of S99, CDA, and FNL. This suggests that grid nudging over fits the RAM small-scale features to the driving global fields. CDA on the other hand produces a distribution comparable to that of S99 at both large- and small-scales that is more consistent with the FNL fields.

To further assess the significance of the larger variability in the similarity index resulting from spectral nudging and CDA, we performed a linear regression analysis between the differences of WRF outputs and NCEP/NCAR fields, and those of FNL to NCEP/NCAR fields. The corresponding scatter plots are presented in Figures 3 and 4 for T500 (Fig. 3) and KE500 (Fig. 4). CDA and spectral nudging improve the correlation and slope with FNL compared to grid nudging at both large- and small-scales. Similar results are obtained at 850 hPa (Fig. 5 and 6). Overall, the differences distributions analysis results are consistent with those obtained from the analysis of the similarity index above, indicating the relevance of the small differences in the similarity index in evaluating the performances of the downscaling techniques, as has been already suggested by Lui et al. (2012).

The differences in KE similarities computed between NCEP/NCAR and WRF using spectral nudging and CDA may reflect significant differences in the surface sea level pressure (SLP), relative humidity (RH) at 850 hPa and 500 hPa, clouds, and precipitation patterns of the downscaled solutions, which are important parameters for regional climate, air quality, and hydrological modeling studies. We therefore analyzed the mean SLP (Fig. 7) and RH at 850 hPa (Fig. 8) and 500 hPa (Fig. 9) as simulated by the different downscaling methods along with their corresponding NCEP/NCAR fields. The dominant pressure patterns observed in the NCEP/NCAR (Fig. 7a) are a relatively low pressure about 1010 hPa to 1015 hPa between 15N to 15S and a high pressure above 1020 hPa north of 15N. Similar patterns are well produced by the downscaling simulations, but not the control run (Fig. 7c), which shows a larger bias of about 20-25 hPa over the oceanic regions and of about 20 hPa over the Arabian Peninsula. The spatial patterns of RH at 850 hPa from NCEP/NCAR clearly indicate (Fig. 8) the dry conditions over the desert regions (around 15N to 25N) and the coastal regions of the Arabian Sea between the Somali coast and northwestern India. These dry and humid conditions are well simulated by CDA and spectral nudging. The control and grid-nudging simulations (Fig. 8c and 8d) show higher humidity over the dry regions and mountains compared to NCEP/NCAR. Similar to 850 hPa, the RH patterns (Fig. 9) at 500 hPa level are also well simulated by all experiments, except the control run. Low humid values north of the equator extending up to 30N, and high humid values south of the equator extending up to 20S are dominant features in the NCEP/NCAR (Fig. 9a). Again, these are well reproduced by all simulations except the control run (Fig. 9c), which exhibits low humid values around the equator from 20S to 30N. The small isolated regions around 10S with higher humidity in CDA (Fig. 9b) and spectral nudging (Fig. 9e, f) are associated with convective activities in the Intertropical Convergence Zone. The changes in the MSLP and RH between the different simulations can modulate the cloud formation and associated rainfall. To confirm this, we compared the total cloud coverage in the different simulations to the Modern-Era Retrospective analysis for Research and Applications data (MERRA, Rienecker et al., 2011) in Figure 10. The mean total cloud coverage from MERRA shows (Fig.10a) a higher coverage of about 70 - 80% between 15S and the equator compared to other regions where the cloud coverage is relatively low (about 10-40%). These observed mean cloud features are well produced by the CDA (Fig.10b) and the spectral nudging simulations (Fig. 10e and f). The control (Fig. 10c) and grid-nudging (Fig. 10d) simulations failed to reproduce the observed cloud patterns as observed by MERRA, with higher cloud coverage over the ocean regions, and over land around the equator. The CDA and S99 solutions are in good agreement with MERRA not only in terms of spatial distribution, but also in terms of magnitude. S99 shows (Fig. 10e) better mean cloud patterns compared to S33, confirming the sensitivity of spectral nudging to the choice of the cut-off wave number.

Given the considerable differences in surface pressure, relative humidity, cloud coverage, and KE similarities with the different nudging methods, investigating the associated changes in the simulation of precipitation is of interest. We therefore compared the spatial patterns of the total rainfall from the different WRF runs with and without nudging to the Global Precipitation Measurements (GPM, Huffman, 2016). Daily GPM observations are extracted for the studied domain from half-hourly data at $10 \times 10$ km. The GPM observations suggest (Fig. 11a) isolated rainfall patterns over the AP and adjoining regions; heavy rainfall over the northern parts; small localized mountain regions with isolated heavy rainfall between the equator to south of the AP; and a continuous rain belt south of the equator. The control (Fig. 11c) and grid nudging (Fig. 11d) simulations fail to predict both rainfall intensity and spatial distribution over the study region. The observed large-scale features and mesoscale variations are much better reproduced by CDA (Fig. 11b) and spectral nudging (Fig. 11e, f). Experiments with different wave numbers suggests that the higher wave numbers (S99) generate a comparable precipitation patterns to those of GMP, primarily over northern Saudi Arabia, which are comparable to those simulated by CDA.

The qualitative analysis of the spatial distribution of the different atmospheric variables suggest that both CDA and S99 are in good agreement with the observations. For a quantitative analysis, and to assess the added value (AV) from the different nudging methods with respect to the FNL observations compared to the control run, we computed the modified Brier skill score (BSS) for T850 and KE850 as representatives of the thermodynamic and dynamic fields and plotted the results in Fig. 12 and 13. CDA and spectral nudging result in important improvement for both T850 (Fig. 12) and K850 (Fig. 13) over the entire model domain compared to the control run, except for a small mountainous region of northeast Africa, southwestern Red Sea, and northeastern Arabian Peninsula. The BSS results are overall consistent with those of the analysis of similarity.

## 5. Summary

Nudging techniques enforce a dynamical balance between Regional Atmospheric Models (RAMs) and global atmospheric fields play an important role in producing reliable regional simulations. In this study, we successfully implemented and tested a new method, continuous data assimilation (CDA), for dynamical downscaling of a three-dimensional general circulation atmospheric model. We also evaluated its performance in reproducing the large- and small-scale features of the downscaled region. We further compared the results of CDA against those obtained using the standard grid and spectral nudging methods. As one of the most widely used atmospheric models for downscaling, we used the Weather Research and Forecast (WRF) model in our downscaling experiments, which offers advanced packages for downscaling with grid and spectral nudging.

WRF model simulations at 25 km x 25 km (~0.25° × 0.25°) horizontal resolution were carried out over a period of 31 days starting on January 1, 2016 with no, grid, spectral, and CDA nudging. We also performed three more simulations with spectral nudging using different wave cut-off numbers (3, 5, and 9) to assess its sensitivity to the choice of retained wave number. The initial conditions and boundary forcing (updated every six hours) were obtained from the NCEP/NCAR reanalysis available at 2.5° × 2.5°. Another independent source of data, the NCEP Final Analysis (FNL) available at 0.25° × 0.25° resolution, was used for validation. The evaluation of the downscaled fields was performed based on the similarity metric suggested by Liu et al. (2012) for two different scales 2000 km and 300 km representing the large- and small-scale, respectively. We analyzed different parameters such as temperature, pressure, winds and kinetic energy to assess the performance of the different downscaling methods. The Brier skill score (BSS) index for temperature and kinetic energy at 850 hPa was also computed to assess the added value (AV) of different nudging methods with respect to FNL.

Overall, the downscaling experiments well captured the features of the driving large-scale fields. At small-scales, spectral nudging with well-tuned cut-off wave number and CDA produced stronger spatial and temporal variability than grid nudging, suggesting the ability of these methods to enable more small-scale features in the model simulations, which may be constrained by coarse downscaled fields in grid nudging. Spectral nudging and CDA performances were also significantly closer to FNL than grid and no-nudging for surface pressure (PS), temperature, and kinetic energy. The choice of the cut-off wave number was very important in spectral nudging: an increase in the number of waves produced better small-scale features, up to a certain wave number after which its results became close to grid nudging. The results of the newly implemented CDA method were consistent with the FNL and its performance is comparable to the best spectral nudging simulation in producing both large- and small-scale features.

Our initial results suggest that the CDA method is a promising approach for dynamical downscaling and enables to efficiently retain the balance between the RAMs and GCMs by constraining the RAM large-scale features to the GCM fields and allowing to develop its own small-scale features without the need of a spectral decomposition, which also means saving in terms of computational time. More tests and experiments will be conducted in future studies to further assess the performance of CDA with different grid resolution, other driving global fields, different interpolation operators, and different physical processes for predicting various regional atmospheric phenomena.

**Acknowledgments**

The research was supported by the office of Sponsor Research (OSR) at King Abdullah University of Science and Technology (KAUST) under the Virtual Red Sea Initiative (Grant #REP/1/3268-01-01) and the Saudi ARAMCO-KAUST Marine Environmental Observatory (SAKMEO). This research made use of the Supercomputing Laboratory resources at KAUST. The work of Edriss S. Titi was supported in part by ONR grant N00014-15-1-2333, the Einstein Stiftung/Foundation - Berlin, through the Einstein Visiting Fellow Program, and by the John Simon Guggenheim Memorial Foundation.

Table 1. Mean and standard deviation of the similarity between the RAM and NCEP/NCAR at large-scales

| Mean of the similarity for large-scale waves (2000 km) | | | | | | | | | |
|---|---|---|---|---|---|---|---|---|---|
| | PS | T850 | U850 | V850 | KE850 | T500 | U500 | V500 | KE500 |
| CDA | 0.99998605 | 0.99999714 | 0.97167140 | 0.97152543 | 0.96723545 | 0.99999964 | 0.99854946 | 0.99049723 | 0.99777579 |
| Grid | 0.99998581 | 0.99999744 | 0.98094010 | 0.97748172 | 0.99035347 | 0.99999976 | 0.99912435 | 0.99274999 | 0.99860662 |
| No-Nudging | 0.99986231 | 0.99972951 | 0.05710086 | 0.33376178 | 0.46544221 | 0.99996454 | 0.84965920 | 0.30325863 | 0.87648469 |
| S33 | 0.99998665 | 0.99999714 | 0.94595301 | 0.91221452 | 0.91652679 | 0.99999917 | 0.99386102 | 0.96675873 | 0.99145013 |
| S55 | 0.99998653 | 0.99999756 | 0.97086024 | 0.96998733 | 0.94466907 | 0.99999964 | 0.99846143 | 0.98823738 | 0.99735242 |
| S99 | 0.99998629 | 0.99999785 | 0.97609609 | 0.97662634 | 0.98445141 | 0.99999964 | 0.99895734 | 0.99211472 | 0.99839503 |
| FNL | 0.99999946 | 0.99999064 | 0.84402716 | 0.82412124 | 0.90915436 | 0.99999577 | 0.97440130 | 0.85004336 | 0.98543674 |
| Standard deviation of the similarity for large-scale waves (2000 km) | | | | | | | | | |
| | PS | T850 | U850 | V850 | KE850 | T500 | U500 | V500 | KE500 |
| CDA | 0.000005 | 0.000001 | 0.014733 | 0.016004 | 0.030686 | 0.000000 | 0.000770 | 0.007707 | 0.001503 |
| Grid | 0.000005 | 0.000001 | 0.010876 | 0.013207 | 0.005458 | 0.000000 | 0.000378 | 0.006230 | 0.000723 |
| No-Nudging | 0.000042 | 0.000087 | 0.466062 | 0.303761 | 0.425571 | 0.000020 | 0.072125 | 0.577311 | 0.061159 |
| S33 | 0.000005 | 0.000001 | 0.028365 | 0.041847 | 0.060352 | 0.000000 | 0.003972 | 0.029892 | 0.004708 |
| S55 | 0.000005 | 0.000001 | 0.014842 | 0.016339 | 0.047274 | 0.000000 | 0.000767 | 0.009983 | 0.001695 |
| S99 | 0.000005 | 0.000001 | 0.013099 | 0.012828 | 0.010660 | 0.000000 | 0.000456 | 0.007248 | 0.000999 |
| FNL | 0.000000 | 0.000003 | 0.086345 | 0.100384 | 0.056032 | 0.000002 | 0.014739 | 0.128554 | 0.008931 |

PS: surface pressure; T850 and T500: Temperature at 850 and 500 hPa levels; U850 and U500: Zonal wind components at 850 and 500 hPa levels; V850 and V500: Meredional wind components at 850 and 500 hPa levels; KE850 and KE500: Kinetic Energy at 850 and 500 hPa levels.

Table 2. Mean and standard deviation of the similarity between the RAM and NCEP/NCAR at small-scales

| | Mean of the similarity for small-scale waves (300 km) | | | | | | | | |
|---|---|---|---|---|---|---|---|---|---|
| | PS | T850 | U850 | V850 | KE850 | T500 | U500 | V500 | KE500 |
| **CDA** | 0.99998337 | 0.99999058 | 0.91240192 | 0.89633244 | 0.89494669 | 0.99999863 | 0.98933464 | 0.97649366 | 0.99176407 |
| **Grid** | 0.99998373 | 0.99999225 | 0.95990425 | 0.94997531 | 0.97067863 | 0.99999911 | 0.99773675 | 0.98945946 | 0.99676031 |
| **No-Nudging** | 0.99985540 | 0.99964535 | -0.0519144 | 0.07106254 | 0.08139410 | 0.99993056 | 0.74900043 | 0.15007074 | 0.74443442 |
| **S33** | 0.99998409 | 0.99998194 | 0.68842328 | 0.63695365 | 0.63966984 | 0.99999440 | 0.94511408 | 0.82956451 | 0.92996812 |
| **S55** | 0.99998403 | 0.99998814 | 0.83709604 | 0.78103155 | 0.79188532 | 0.99999708 | 0.97539669 | 0.92701846 | 0.96858078 |
| **S99** | 0.99998385 | 0.99999207 | 0.92231774 | 0.90175653 | 0.92826903 | 0.99999875 | 0.99454635 | 0.97957581 | 0.99282628 |
| **FNL** | 0.99999863 | 0.99997193 | 0.65930980 | 0.60649222 | 0.70462215 | 0.99998736 | 0.92336237 | 0.72812873 | 0.92406327 |
| | Standard deviation of the similarity for small-scale waves (300 km) | | | | | | | | |
| | PS | T850 | U850 | V850 | KE850 | T500 | U500 | V500 | KE500 |
| **CDA** | 0.000006 | 0.000003 | 0.027247 | 0.034035 | 0.074811 | 0.000000 | 0.004850 | 0.013146 | 0.004275 |
| **Grid** | 0.000005 | 0.000003 | 0.011793 | 0.017309 | 0.010606 | 0.000000 | 0.000569 | 0.005710 | 0.001229 |
| **No-Nudging** | 0.000043 | 0.000104 | 0.413987 | 0.270213 | 0.461699 | 0.000032 | 0.096030 | 0.493374 | 0.107139 |
| **S33** | 0.000005 | 0.000004 | 0.120302 | 0.105091 | 0.173989 | 0.000001 | 0.015521 | 0.097885 | 0.029512 |
| **S55** | 0.000005 | 0.000003 | 0.040479 | 0.058794 | 0.075941 | 0.000001 | 0.007464 | 0.040535 | 0.013222 |
| **S99** | 0.000006 | 0.000003 | 0.018289 | 0.026641 | 0.021350 | 0.000000 | 0.001449 | 0.011168 | 0.003153 |
| **FNL** | 0.000000 | 0.000005 | 0.101990 | 0.104339 | 0.087572 | 0.000003 | 0.024534 | 0.139838 | 0.026483 |

PS: surface pressure; T850 and T500: Temperature at 850 and 500 hPa levels; U850 and U500: Zonal wind components at 850 and 500 hPa levels; V850 and V500: Meredional wind components at 850 and 500 hPa levels; KE850 and KE500: Kinetic Energy at 850 and 500 hPa levels.

**Table 3. Mean and standard deviation of the distribution between the RAM and NCEP at large-scales**

| | Mean of large-scale waves (2000 km) | | | | | |
|---|---|---|---|---|---|---|
| | **T850** | **U850** | **V850** | **T500** | **U500** | **V500** |
| **FNL** | 0.332 | 0.058 | -0.196 | -0.019 | -0.009 | 0.317 |
| **CDA** | -0.204 | -0.100 | 0.058 | -0.054 | 0.027 | 0.045 |
| **Grid** | -0.208 | -0.110 | 0.085 | -0.066 | 0.027 | 0.058 |
| **S99** | -0.164 | -0.101 | 0.051 | -0.054 | 0.042 | 0.054 |
| | Standard deviation of large-scale waves (2000 km) | | | | | |
| | **T850** | **U850** | **V850** | **T500** | **U500** | **V500** |
| **FNL** | 0.225 | 0.394 | 0.281 | 0.161 | 0.507 | 0.474 |
| **CDA** | 0.079 | 0.140 | 0.107 | 0.061 | 0.100 | 0.101 |
| **Grid** | 0.068 | 0.116 | 0.090 | 0.053 | 0.089 | 0.086 |
| **S99** | 0.075 | 0.129 | 0.090 | 0.059 | 0.092 | 0.091 |

T850 and T500: Temperature at 850 and 500 hPa levels; U850 and U500: Zonal wind components at 850 and 500 hPa levels; V850 and V500: Meredional wind components at 850 and 500 hPa levels.

**Table 4. Mean and standard deviation of the distribution between the RAM and NCEP at small-scales**

| | Mean of small-scale waves (300 km) | | | | | |
|---|---|---|---|---|---|---|
| | **T850** | **U850** | **V850** | **T500** | **U500** | **V500** |
| **FNL** | 0.357 | 0.060 | -0.004 | 0.093 | -0.004 | 0.260 |
| **CDA** | -0.194 | -0.046 | -0.043 | 0.042 | -0.043 | 0.024 |
| **Grid** | -0.206 | -0.088 | -0.055 | 0.028 | -0.055 | 0.040 |
| **S99** | -0.155 | -0.064 | 0.029 | -0.044 | 0.045 | 0.036 |
| | Standard deviation of small-scale waves (300 km) | | | | | |
| | **T850** | **U850** | **V850** | **T500** | **U500** | **V500** |
| **FNL** | 0.250 | 0.390 | 0.281 | 0.153 | 0.524 | 0.462 |
| **CDA** | 0.080 | 0.130 | 0.107 | 0.060 | 0.097 | 0.090 |
| **Grid** | 0.075 | 0.108 | 0.090 | 0.052 | 0.093 | 0.078 |
| **S99** | 0.075 | 0.117 | 0.086 | 0.058 | 0.098 | 0.082 |

T850 and T500: Temperature at 850 and 500 hPa levels; U850 and U500: Zonal wind components at 850 and 500 hPa levels; V850 and V500: Meredional wind components at 850 and 500 hPa levels.

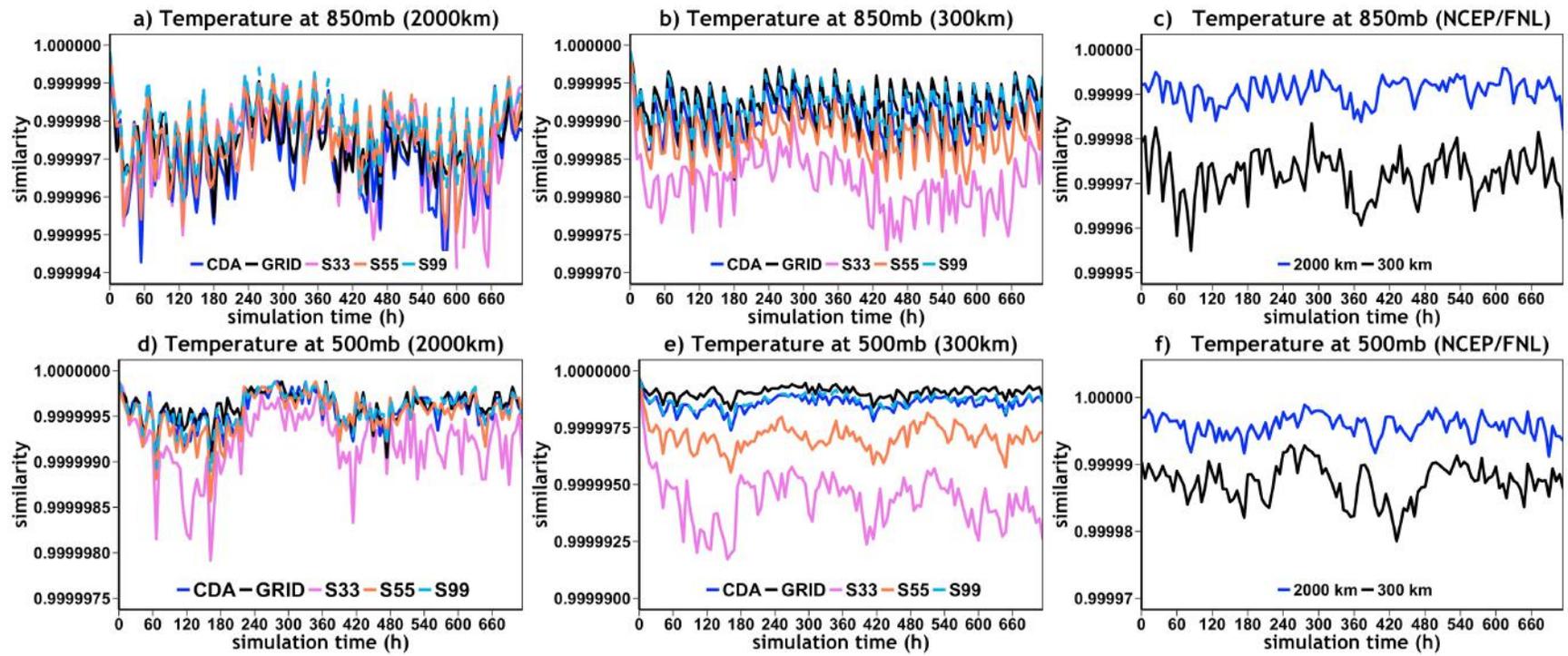

Figure 1. Time series of similarity in temperature at 850 (top panel) and 500 hPa (bottom panel) between NCEP and different experiments (a, b, d, and e) with the RAM model, and between NCEP and FNL (c and f) at large- and small-scale waves.

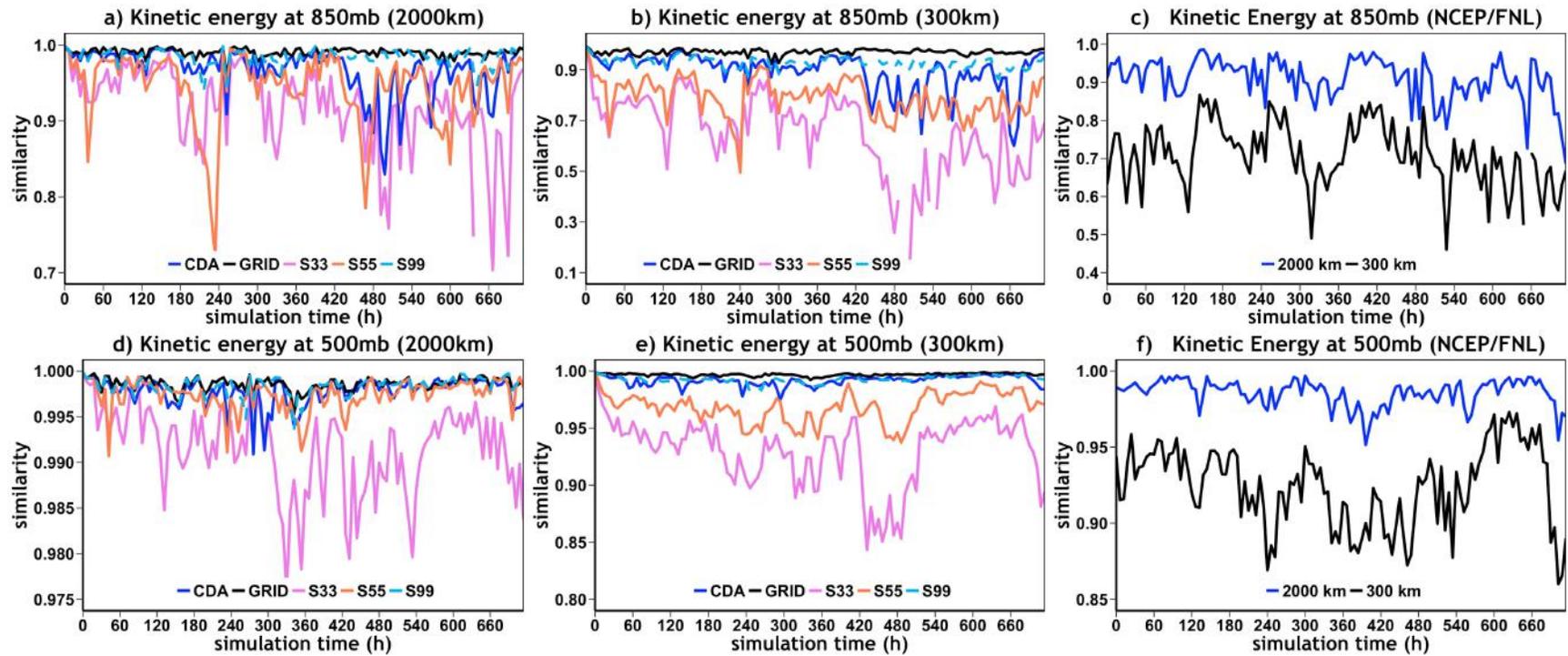

Figure 2. Time series of similarity in kinetic energy at 850 (top panel) and 500 hPa (bottom panel) between NCEP and different experiments (a, b, d, and e) with the RAM model, and between NCEP and FNL (c and f) at large- and small-scale waves.

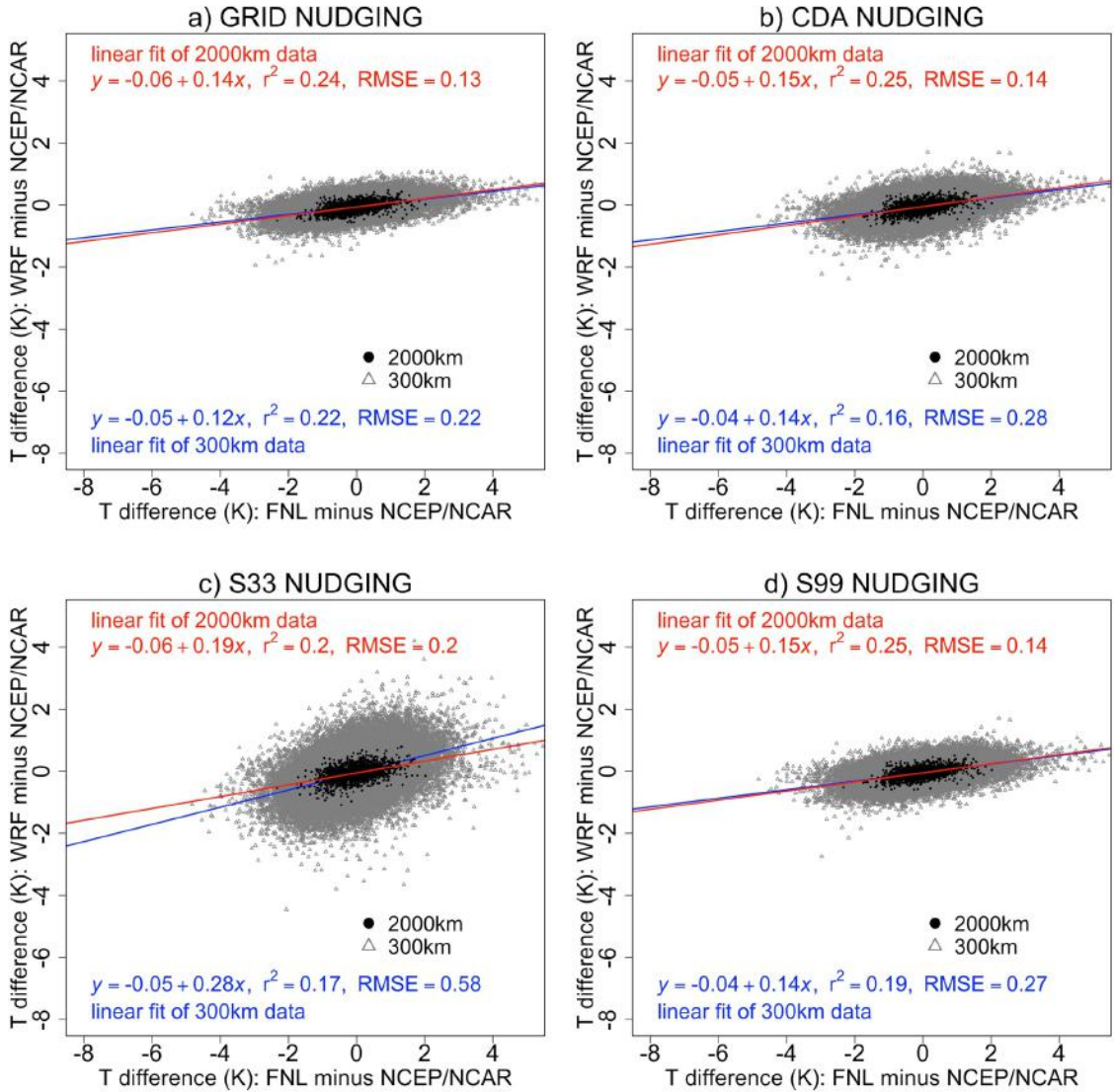

Figure 3. Correlations between the temperature anomalies from NCEP/NCAR at 500 hPa for different WRF model experiments of with a) Grid nudging, b) CDA, c) Spectral nudging with three waves, and d) Spectral nudging with nine waves in both X and Y-directions for large- and small-scale waves.

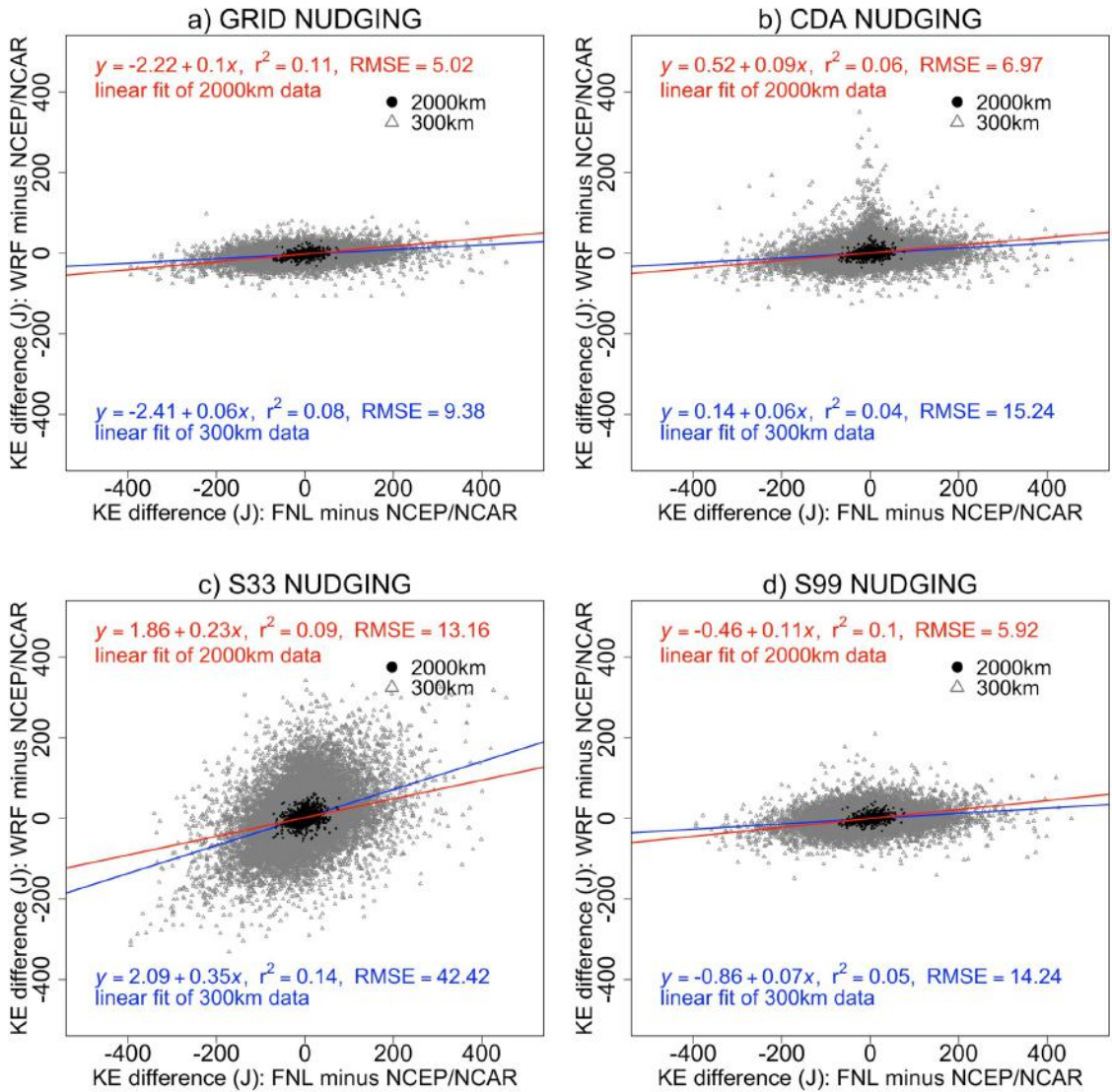

Figure 4. Correlations between kinetic energy anomalies from NCEP/NCAR at 500 hPa for different WRF model experiments of with a) Grid nudging, b) CDA, c) Spectral nudging with three waves, and d) Spectral nudging with nine waves in both X- and Y-directions for large- and small-scale waves.

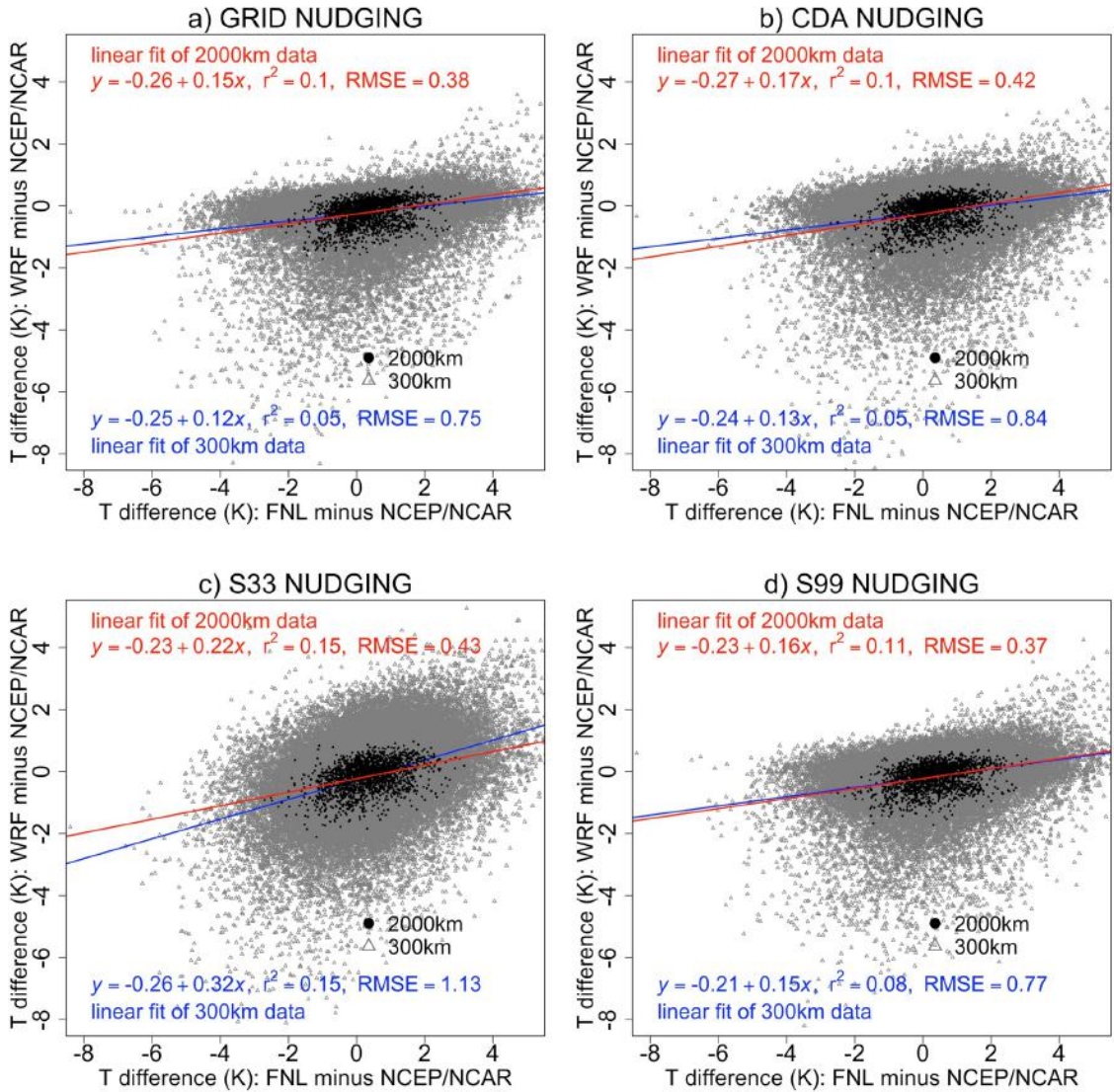

Figure 5. Correlations between the temperature anomalies from NCEP/NCAR at 850 hPa for different WRF model experiments of with a) Grid nudging, b) CDA, c) Spectral nudging with three waves, and d) Spectral nudging with nine waves in both X and Y-directions for large- and small-scale waves.

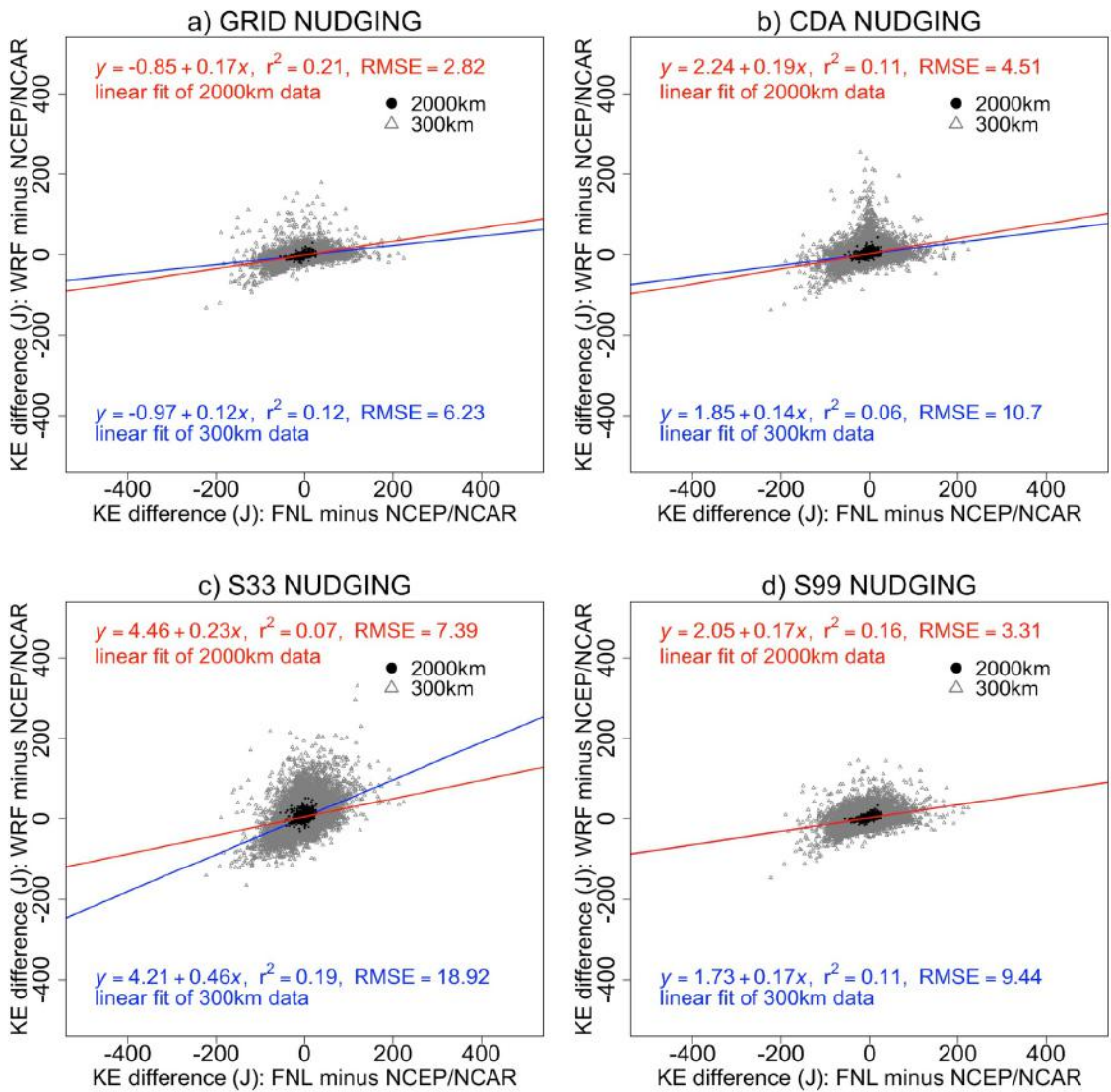

Figure 6. Correlations between kinetic energy anomalies from NCEP/NCAR at 850 hPa for different WRF model experiments of with a) Grid nudging, b) CDA, c) Spectral nudging with three waves, and d) Spectral nudging with nine waves in both X- and Y- directions for large- and small-scale waves.

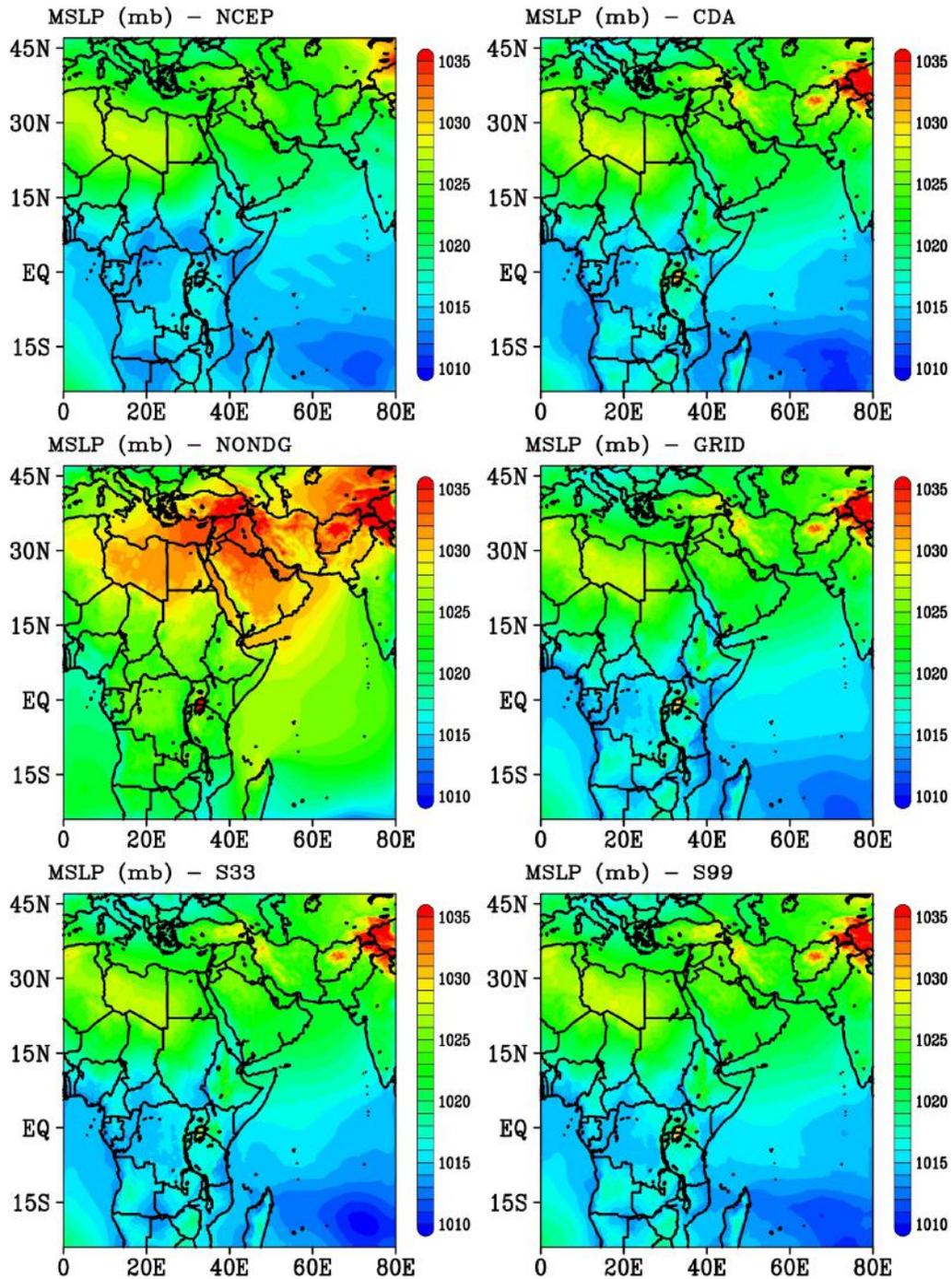

Figure 7. Spatial distribution of mean surface pressure (hPa) from a) NCEP/NCAR and different WRF model simulations of with b) CDA, c) No-nudging (NONDG), d) Grid-nudging (GRID), e) Spectral nudging with three waves (S33), and f) Spectral nudging ,with nine waves (S99) in both X and Y- directions for January 1–31, 2016.

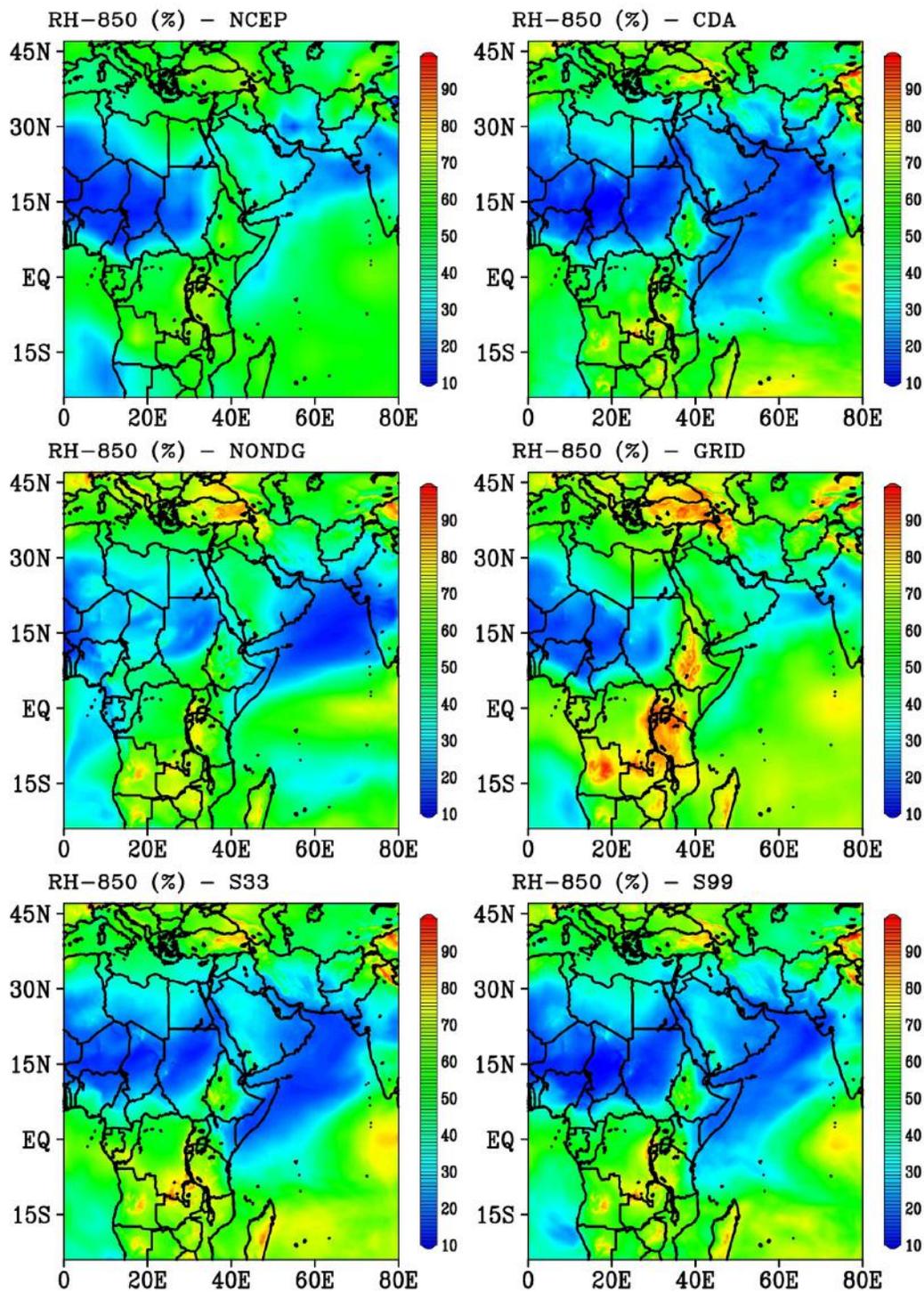

Figure 8. Spatial distribution of Relative humidity (%) at 850 hPa from a) NCEP/NCAR and different WRF model simulations of with b) CDA, c) No-nudging (NONDG), d) Grid-nudging (GRID), e) Spectral nudging with three waves (S33), and f) Spectral nudging with nine waves (S99) in both X and Y- directions for January 1–31, 2016.

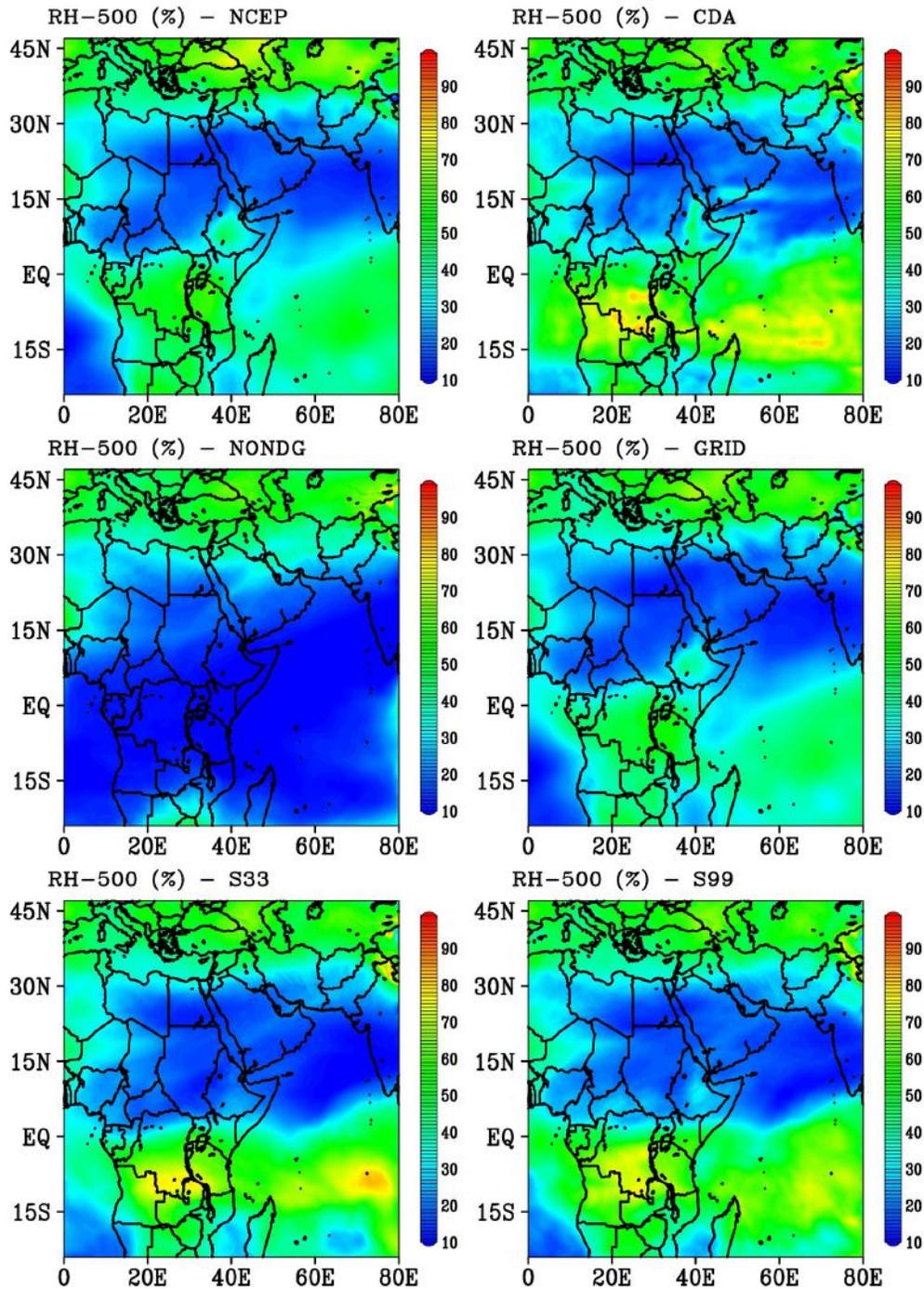

Figure 9. Spatial distribution of Relative humidity (%) at 500 hPa from a) NCEP/NCAR and different WRF model simulations of with b) CDA, c) No-nudging (NONDG), d) Grid-nudging (GRID), e) Spectral nudging with three waves (S33), and f) Spectral nudging with nine waves (S99) in both X and Y- directions for January 1–31, 2016.

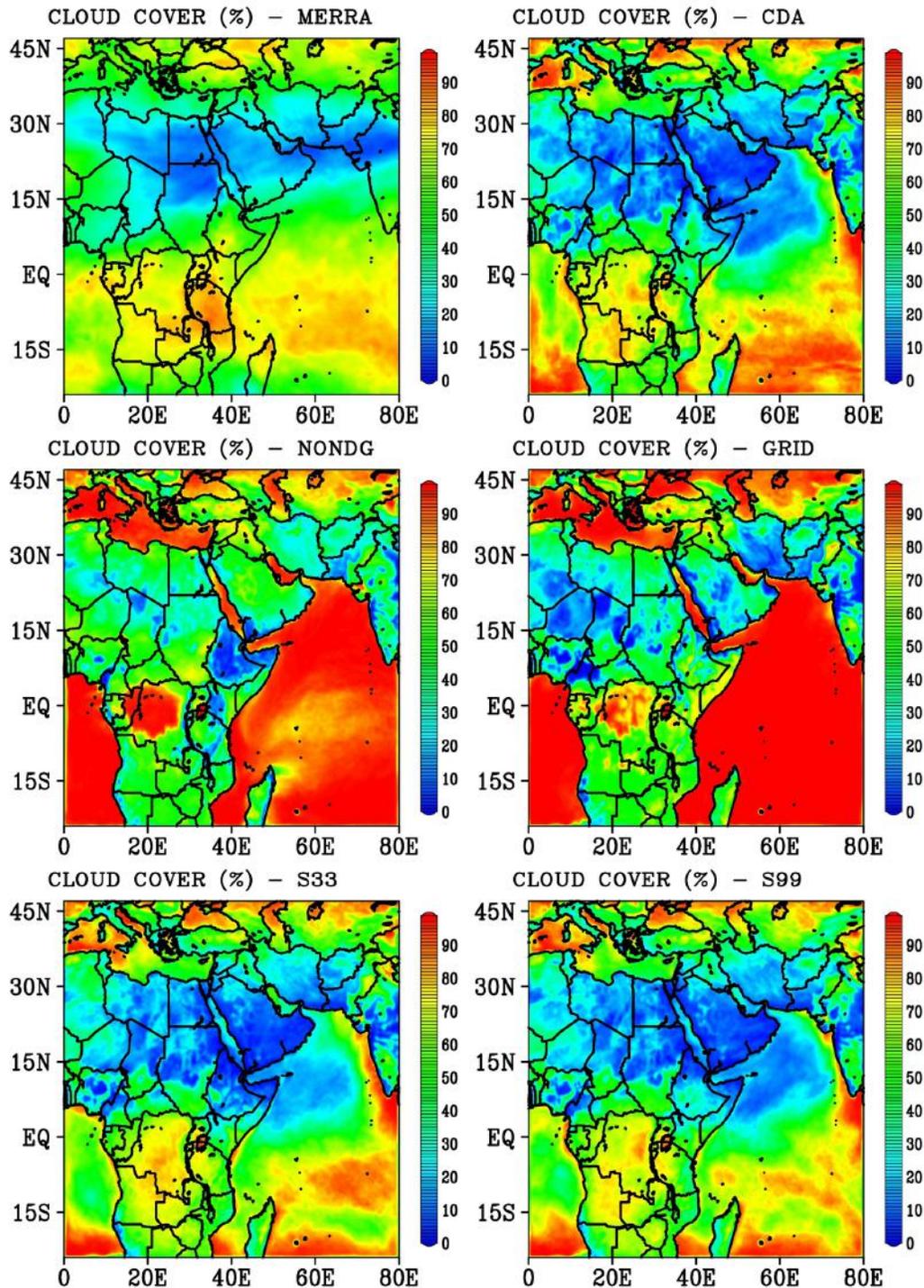

Figure 10. Spatial distribution of total cloud cover (%) from a) MERRA observations and different WRF model simulations of with b) CDA, c) No-nudging (NONDG), d) Grid-nudging (GRID), e) Spectral nudging with three waves (S33), and f) Spectral nudging with nine waves (S99) in both X and Y- directions for January 1–31, 2016.

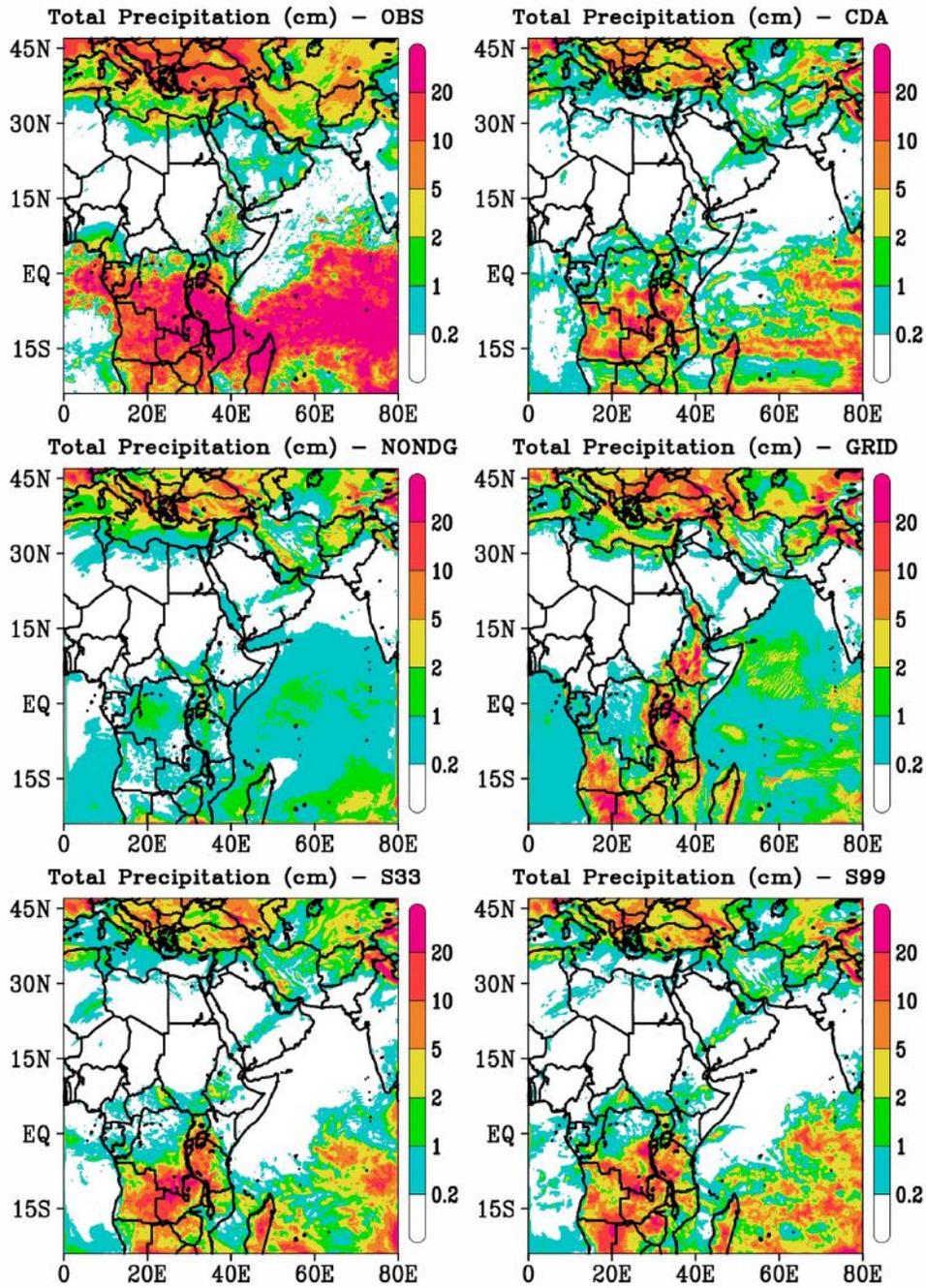

Figure 11. Spatial distribution of total precipitation (mm) from a) GPM observations (OBS) and different WRF model simulations of with b) CDA, c) No-nudging (NONDG), d) Grid-nudging (GRID), e) Spectral nudging with three waves (S33), and f) Spectral nudging with nine waves (S99) in both X and Y- directions for January 1–31, 2016.

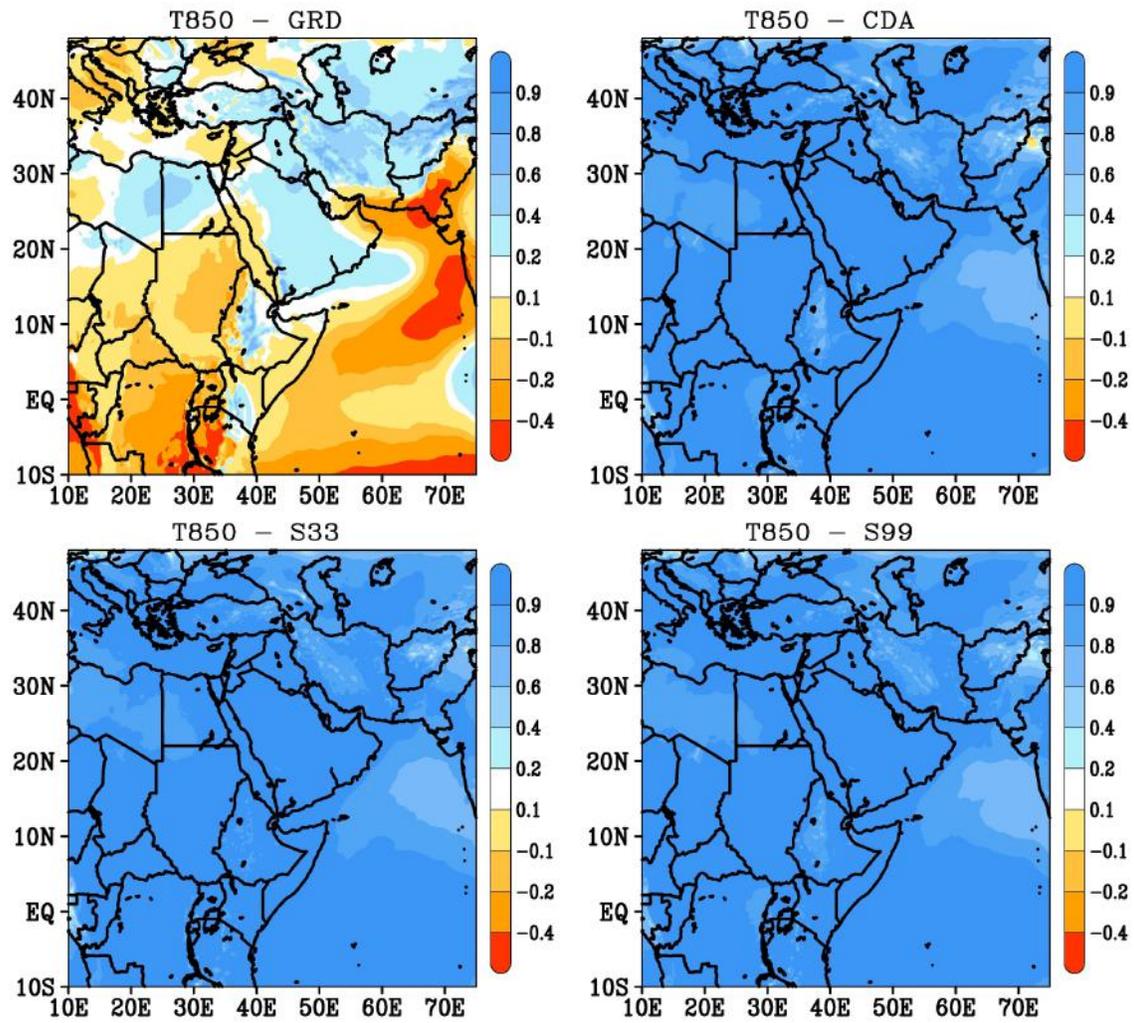

Figure 12. Brier skill score distribution of the different nudging experiments relative to the control simulations with FNL temperature (C) at 850 hPa level as a reference (true) field.

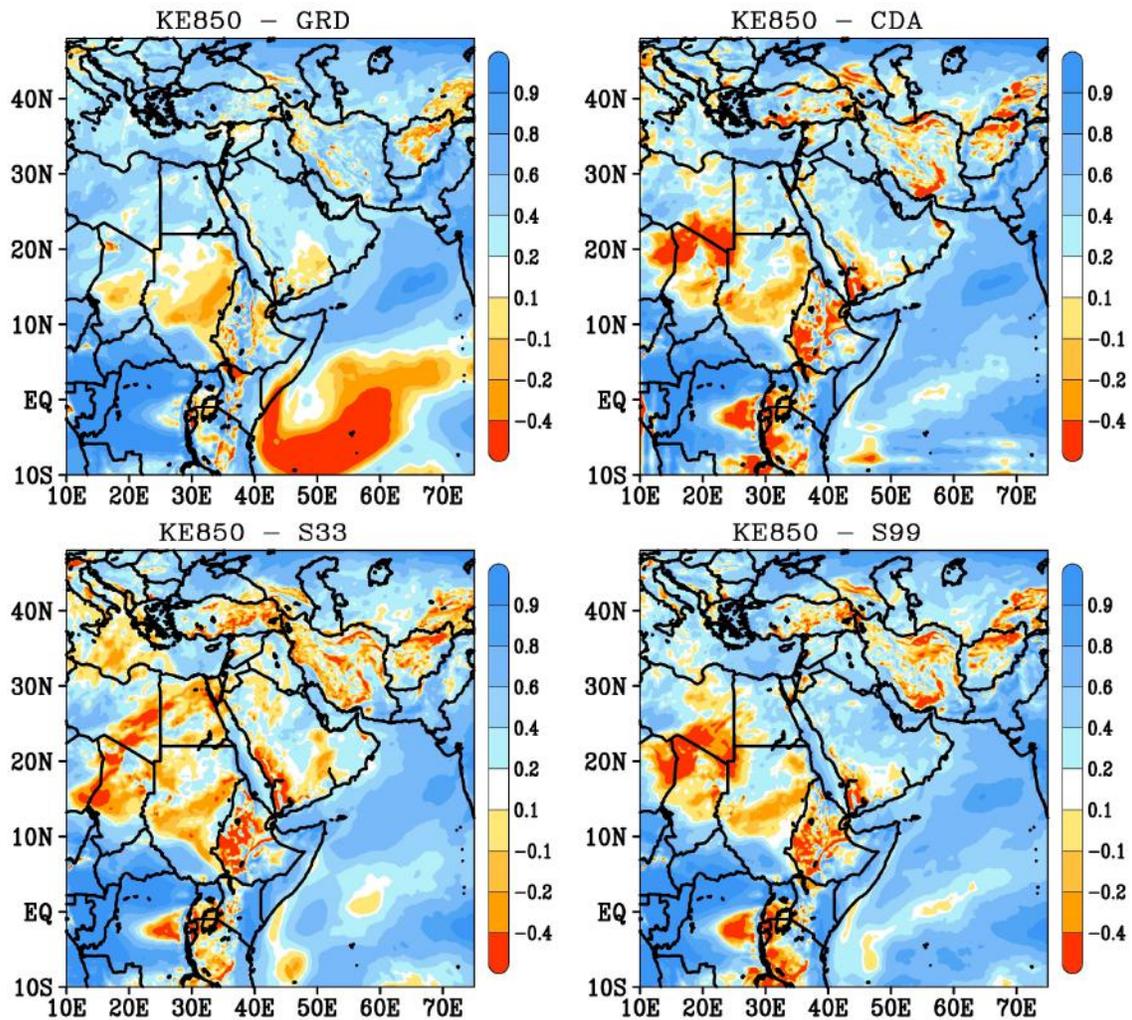

Figure 13. Brier skill score distribution of the different nudging experiments relative to the control simulations with FNL kinetic energy (Joule) at 850 hPa level as a reference (true) field.